\documentclass[%
 reprint,
 amsmath,amssymb,
 aps,
showkeys]{revtex4-2}

\usepackage{graphicx}
\usepackage{dcolumn}
\usepackage{bm}
\usepackage{hyperref}
\hypersetup{colorlinks=true, linkcolor=blue, citecolor=blue}

\begin{document}

\preprint{APS/123-QED}

\title{Energy-Dependent Magnetic Modifications in HOPG via Microbeam Scanning }

\author{Ram Kumar}
\email{ramkumar@iitk.ac.in}
\author{Aditya H. Kelkar}
\email{akelkar@iitk.ac.in}
\affiliation{Department of Physics, Indian Institute of Technology Kanpur, Uttar Pradesh 208016, India}

\author{Neeraj Shukla}
\author{Paras Poswal}
\author{Sheshamani Singh}
\affiliation{Department of Physics, National Institute of Technology Patna, Bihar 800005, India}

\begin{abstract}
Medium-energy ion irradiation is a promising technique for inducing magnetism in materials with partially filled d or f electron bands. This approach enables precise control over the density and spatial distribution of irradiation-induced defects, which play a crucial role in modifying the electronic and magnetic properties of the system. The primary objective of this experiment was to investigate the influence of ion energy variation on the magnetic properties of highly oriented pyrolytic graphite (HOPG). To achieve this, HOPG samples were irradiated with protons (1–3 MeV) and carbon ions (600 keV–2 MeV). A significant change in the magnetic moment was observed with respect to the irradiation energy for both ion species. The effect of energy variation was analyzed using a vibrating sample magnetometer (VSM) and SRIM simulations. The results demonstrate that ion-beam–induced magnetic ordering strongly depends on both the ion species and the beam energy. Magnetic measurements were performed with varying irradiation energies, showing that carbon ion irradiation produces a higher degree of magnetic ordering compared to proton irradiation at the same dose. The maximum magnetization was obtained at 1.2 MeV carbon ion irradiation. SRIM simulations confirm that carbon ions create a greater number of lattice defects than proton ions, which correlates with the enhanced magnetic response.
\end{abstract}

\maketitle
\section{Introduction:-}

In recent years, considerable attention has been devoted to understanding magnetism in materials that do not possess partially filled d- or f-electron bands.
The study of magnetism in carbon based materials, including graphite and fullerenes, has garnered significant scientific interest over the past several
years\cite{Allemand1991,Coey2002,Mathew2007,Lungu2021, telkhozhayeva2024,walker2016} Given its broad relevance across fields such as electronics and biology, understanding the optimal conditions for inducing magnetism in carbon-based materials is crucial. Since the pioneering discovery of ion beam-induced magnetism by Esquinazi et al., the 
phenomenon has attracted growing research interest \cite{Esquinazi2003}. This discovery has led to a surge of research activity in the field of ion beam-induced 
ferromagnetic ordering. For instance, Amit et al. employed 250 keV Ar and 92 MeV Si ions to irradiate fullerene thin films, comparing the effects of electronic excitation and collisional cascades on the resulting magnetization\cite{Kumar2006}. Similarly, Makarova et al. conducted a comparative study on graphite, 
utilizing proton and helium ion irradiation to investigate the origin of defect-induced magnetism in graphite \cite{Makarova2010}. Proton and helium ion irradiation at varying ion fluxes can generate isolated vacancies in graphite, with densities up to eight times higher compared to other conditions, thereby inducing magnetic ordering at medium ion energies \cite{Makarova2011}. Shukla et al. also investigated the magnetic behavior of highly oriented pyrolytic graphite (HOPG) under carbon ion irradiation and reported that the induced magnetic ordering is strongly dependent on the specific nature of the irradiating ion \cite{Shukla2012}.  Several studies have demonstrated that ion irradiation using species such as protons, helium, silicon, argon, nitrogen, and carbon—under optimized energy and fluence conditions can induce ferromagnetic ordering in carbon-based materials \cite{Kumar2006,Makarova2010,Makarova2011,Shukla2012,Wang2014,Kumar2022a,Stefania2022,Talapatra2006}. Most previous studies have conducted ion irradiation experiments at fixed beam energies. However, our analysis indicates that the energy of the ion beam plays a critical role in determining the extent of induced ferromagnetic ordering. The motivation for the present study arises from the observations of various research groups, which employed different ion energy regimes and reported significantly varying levels of magnetization \cite{Mathew2012,Kumar2022b,KUMAR2023a}.  It has been suggested that ion irradiation induces a variety of structural defects in the target material, and the nature, density, and distribution of these defects vary significantly with ion energy, particularly across different depths\cite{WANG2022}. Esquinazi and co-workers have demonstrated that in situ annealing effects caused by the ion beam itself can suppress or even eliminate the induced magnetic ordering. To mitigate this, they employed a low-temperature bath at the sample stage during irradiation to preserve the defect-induced magnetism\cite{Esquinazi2002}. In this study, we aim to investigate the role of ion beam energy in inducing magnetic ordering by systematically varying the energy of carbon and proton irradiation in highly oriented pyrolytic graphite (HOPG). The results provide insights into how different energy regimes influence defect formation and the resulting ferromagnetic behavior. 

The present study aims to compare the magnetic properties of highly oriented pyrolytic graphite (HOPG) samples irradiated with similar ion doses at varying energies. We report magnetic measurements of HOPG bombarded with both protons and carbon ions, where each set was irradiated at different ion energies while maintaining a constant fluence. Our results reveal that medium-energy ion irradiation leads to the strongest magnetic ordering in HOPG. Post-irradiation, the samples exhibit clear signatures of ferromagnetic behavior, indicating that the energy of the incident ions plays a crucial role in defect-mediated magnetism.

\section{Sample Preparation and microbeam scanning }  
Highly oriented pyrolytic graphite (HOPG) samples of grade 1 quality, characterized by an angular mosaic spread of less than 1°, were procured from SPI Supplies. According to the manufacturer’s specifications, the total impurity concentration was below 10 ppm. For irradiation, flakes with dimensions of 3.0 mm {$×$} 1.0 mm {$× $}0.1 mm were cleaved and affixed to thermally oxidized silicon substrates  using GE varnish to ensure mechanical stability during beam exposure.

Ion irradiation experiments were carried out at the Ion Beam Complex for Science, Engineering and Technology, IIT Kanpur, using a 1.7 MV Tandetron accelerator system (High Voltage Engineering, Europa B.V.) equipped with a dedicated microbeam line. The beamline features a quadrupole triplet electromagnetic lens for precise focusing of low-Z ion species (Z {$<$} 6). Protons were extracted from hydrogen plasma and delivered to the samples at energies ranging from 1 to 3 MeV. Carbon ions were similarly used in the energy range of 600 keV to 1.2 MeV. In both cases, the ion fluence was kept constant at 13 pC/µm² to isolate the effect of ion energy on magnetic behavior. This approach enabled a systematic investigation of the energy-dependent magnetic response of HOPG for different ion species.

The focused ion beam was raster-scanned over a 1.0 mm × 1.0 mm area of the HOPG surface using an electromagnetic scanner controlled by a National Instruments NI-6259 data acquisition card. To ensure full coverage of the sample, this scanning process was repeated across adjacent regions.

Stringent protocols were followed during sample preparation, transfer, and irradiation to avoid any extrinsic contamination. To assess any possible transition metal contamination, particle-induced X-ray emission (PIXE) analysis was performed post-irradiation. The results confirmed that iron concentration in the final samples remained below 20 ppm—slightly above the supplier specification but within acceptable limits for magnetic characterization.

Magnetic measurements were carried out using a vibrating sample magnetometer (ADE Magnetics). To verify reproducibility, a second HOPG sample of the same thickness, irradiated under identical dose and energy conditions, was also characterized.

\section{Results} 
\subsection{Magnetization studies of unirradiated and microbeam irradiated HOPG samples}

\begin{figure}[ht]
\centering
\includegraphics[width=0.9\linewidth]{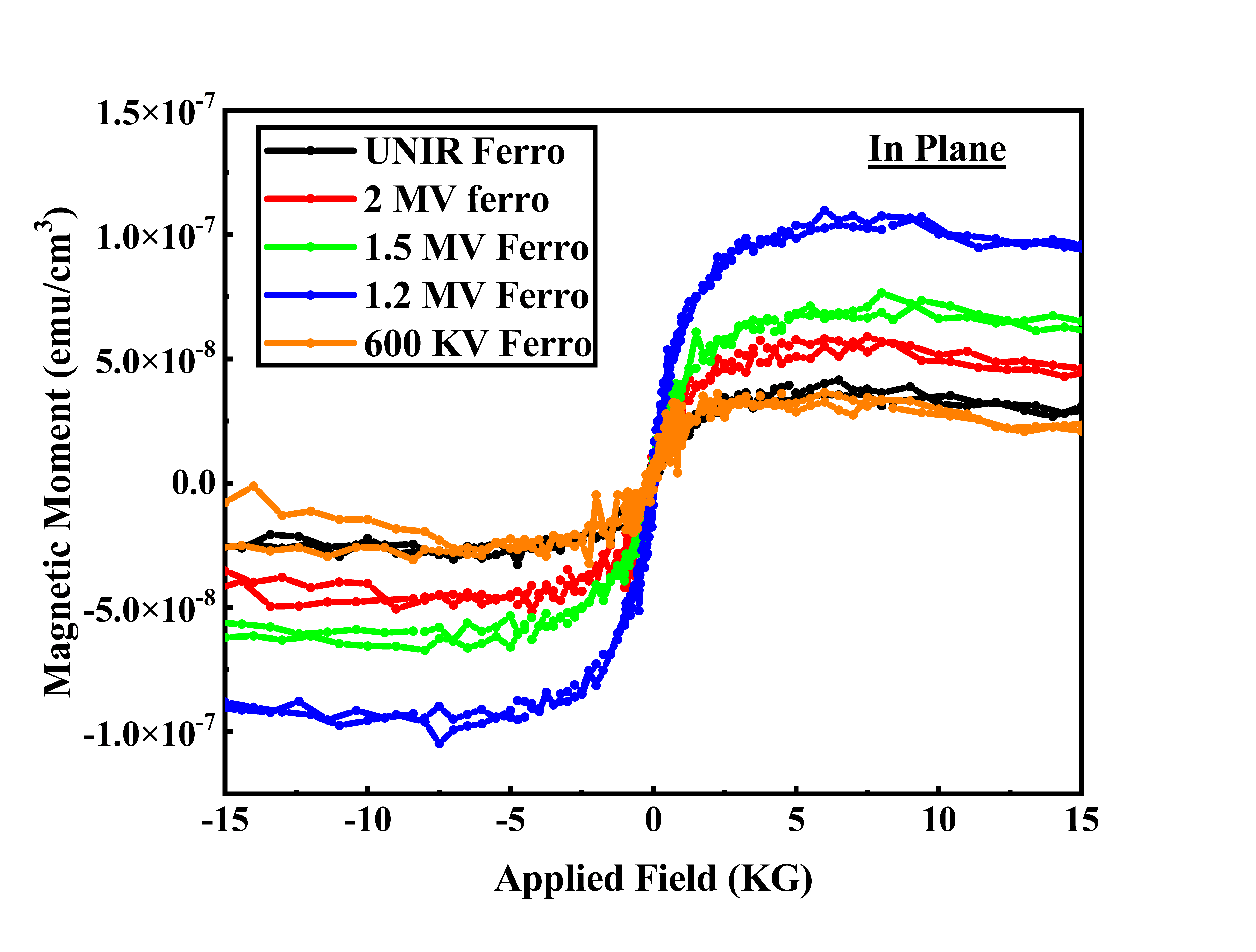}
\caption{\textit{Magnetization response of carbon ion beam irradiated samples at same dose ($100~\text{pC}/\mu\text{m}^2$) with energy variation from 2-0.6 MV
.}}
\label{Figure1.jpg}
\end{figure}

\begin{figure}[ht]
\centering
\includegraphics[width=0.9\linewidth]{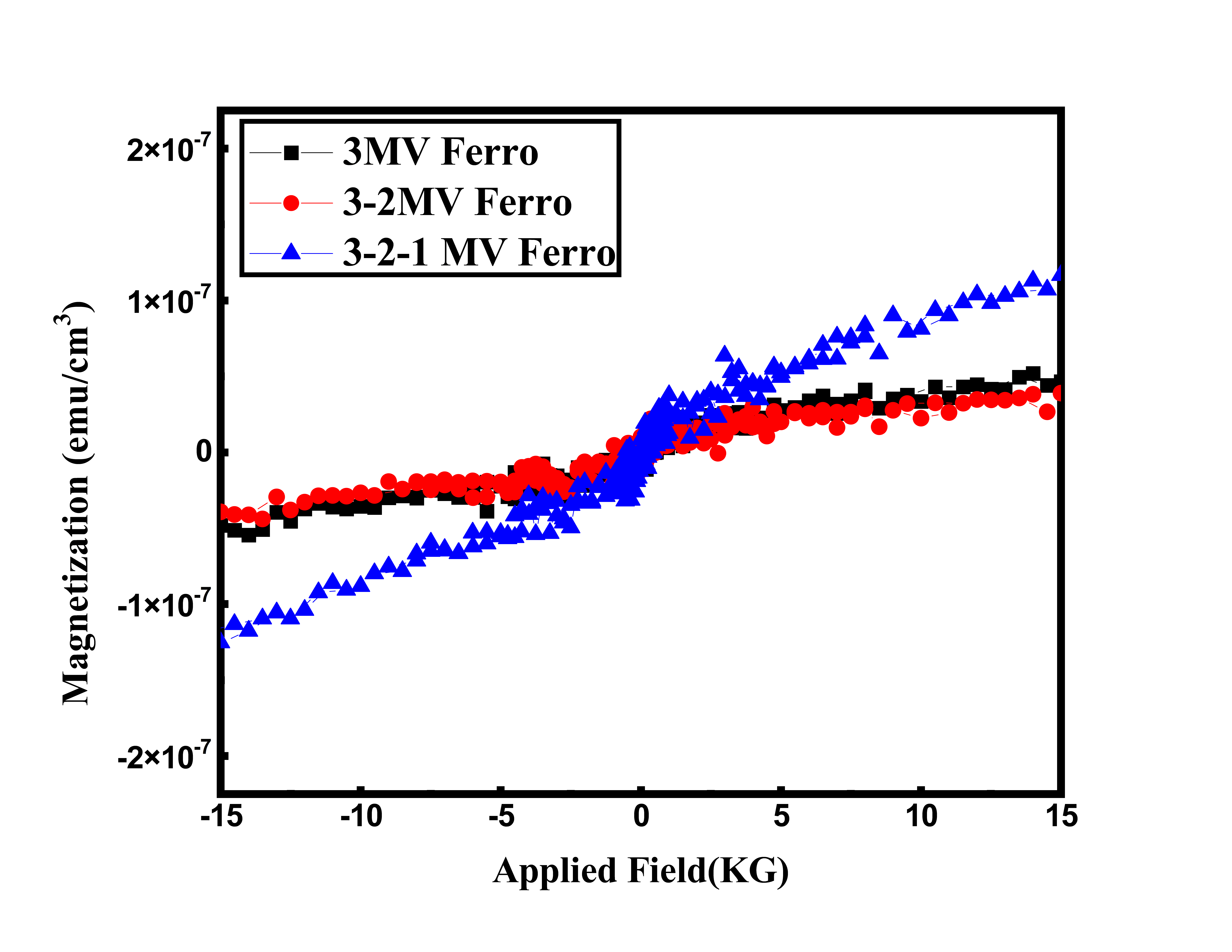}
\caption{\textit{Magnetization response of proton ion beam irradiated samples at same dose ($100~\text{pC}/\mu\text{m}^2$) with energy variation from 3-1 MV
.}}
\label{Figure2.jpg}
\end{figure}

The influence of carbon ion microbeam irradiation, performed in the energy range of 0.6–2.0 MeV at appropriate fluences, on the magnetic behavior of highly oriented pyrolytic graphite (HOPG) flakes is presented in Figure \ref{Figure1.jpg}. For comparison, Figure \ref{Figure2.jpg} shows the corresponding magnetic response of HOPG subjected to proton microbeam scanning in the energy range of 1–3 MeV. These results reveal that both ion species and irradiation energy critically govern the evolution of defect-induced magnetism in HOPG. The pristine sample exhibits a distinct diamagnetic response, as indicated by the consistently negative slope of magnetization.

To determine the optimal irradiation conditions, four sets of HOPG flakes with comparable dimensions were scanned using carbon ions at energies between 0.6 and 2.0 MeV, while three samples were irradiated with protons at energies from 1 to 3 MeV. As shown in Figures \ref{Figure1.jpg} and \ref{Figure2.jpg}, the magnetic moment at 0.6 MeV is $2.5 \times 10^{-8}  \text{emu/cm}^3$, which is nearly identical to that of the pristine sample. The maximum magnetic moment of $9.6 \times 10^{-8}  \text{emu/cm}^3$ is observed for carbon ion irradiation at 1.2 MeV, signifying an optimal energy for defect-mediated magnetic ordering. In contrast, proton irradiation yields the highest magnetic moment of $1.0 \times 10^{-7}  \text{emu/cm}^3$ at 3 MeV.

The pronounced enhancement in magnetic ordering at 1.2 MeV for carbon irradiation is attributed to a balanced interplay between defect generation and ion penetration depth. At this intermediate energy, the defect density within the magnetically active region becomes sufficient to promote long-range magnetic coupling without inducing excessive lattice damage or energy loss beyond the effective range. Conversely, low-energy irradiation results in shallow defect formation, while high-energy ions penetrate deeper, distributing energy over a larger volume and potentially suppressing magnetic interactions due to defect over-saturation or subsurface disorder. At higher energies (1.5–2.0 MeV), the saturation magnetization decreases monotonically, indicating that defect-induced ferromagnetism in HOPG is optimized within a specific energy window of carbon ion microbeam irradiation.


\section{Estimated Defect Density, Penetration Depth, and Energy Dissipation of Ions in HOPG (via SRIM) }

The Stopping and Range of Ions in Matter (SRIM) code was employed to calculate the electronic and nuclear energy losses for each ion type and corresponding energies used in the experiment, as summarized in Table \ref{table2} and \ref{table1}. The selected ion energies span a range that effectively captures variations in both energy loss mechanisms, thereby facilitating a comprehensive understanding of energy-dependent ion–solid interactions in HOPG.

According to the SRIM simulation, the nuclear energy loss dominates at lower ion energies, whereas electronic energy loss becomes more significant at higher energies \cite{Ziegler1985,Ziegler2010}. This transition reflects the fundamental interaction mechanisms governing ion penetration. Consequently, surface damage is more pronounced at low energies due to enhanced sputtering and collision cascades, whereas high-energy ions deposit more energy near the end of their range, potentially inducing subsurface damage.

These trends, presented in Tables \ref{table2} and \ref{table1}, provide crucial insight into how energy loss mechanisms influence defect formation and magnetic property modifications in ion-irradiated HOPG samples.

\begin{table}[ht]
\tiny
\centering
\caption{ \textit{Estimated values of crystallite size, dislocation density, interplanar spacing, and microstrain using XRD data of pristine and irradiated samples}}. 
\begin{tabular}{|c|c|c|c|c|c|}
\hline
Proton   & Nuclear   &  Electronic  & Projeted   &  Longitudinal   & Lateral   \\

Energy& Energy Loss & Energy  Loss & Range & Straggling & Straggling\\
MV& eV/A & eV/A  & $\mu$m & A& A
 \\
\hline
0.6 & 1.447 &8.594 $\times$ $10^{1}$& 0.96 & 766 & 989  \\
\hline
1.2 & 0.836 & 1.309 $\times$ $10^{2}$ & 1.54 & 910 & 1182 \\
\hline
1.5 & 0.697 & 1.309 $\times$ $10^{2}$ & 1.78 & 955 & 1250 \\
\hline
2.0 & 0.551 &1.309 $\times$ $10^{2}$ & 2.16 & 1023 & 1338 \\
\hline

\end{tabular}
\label{table2}
\end{table}

\begin{table}[ht]
\tiny
\centering
\caption{ \textit{Estimated values of crystallite size, dislocation density, interplanar spacing, and microstrain using XRD data of pristine and irradiated samples}}. 
\label{table1}

\begin{tabular}{|c|c|c|c|c|c|}
\hline
Proton   & Nuclear   &  Electronic  & Projeted   &  Longitudinal   & Lateral   \\

Energy& Energy Loss & Energy  Loss & Range & Straggling & Straggling\\
MV& eV/A & eV/A  & $\mu$m & A& A
 \\
\hline
1.0 & 3.432 $\times$ $10^{-3}$ & 5.213& 12.29 & 5295 & 4463  \\
\hline
2.0 & 1.872 $\times$ $10^{-3}$ & 3.209 & 37.45 & 14700 & 12000 \\
\hline
3.0 & 1.309 $\times$ $10^{-3}$ & 2.379 & 47.04 & 30800 & 22600 \\
\hline

\end{tabular}

\end{table}

\section{Impurity Analysis by PIXE (Particle Induced X-ray Emission) }

When unexpected magnetic behavior is observed, the most immediate consideration is the possible influence of magnetic impurities. However, isolated or sparsely distributed impurities—unless present as large aggregates with suitable particle sizes are unlikely to produce stable or measurable magnetism. Nevertheless, it is essential to eliminate any such possibility before drawing conclusions about the intrinsic origin of the observed magnetism.

\begin{figure}[ht]
\centering
\includegraphics[width=0.9\linewidth]{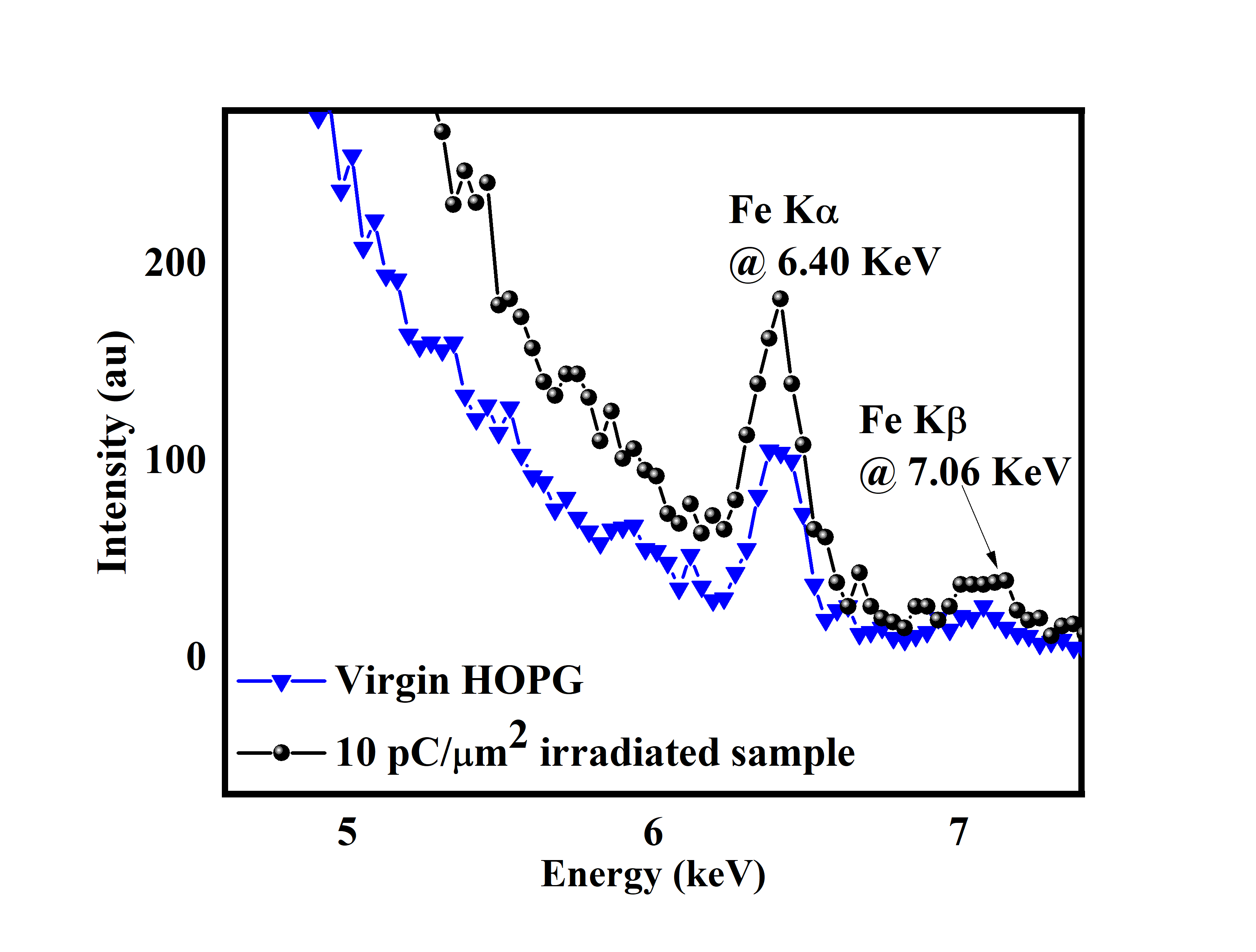}
\caption{\textit{Particle Induced X-ray Emission (PIXE) data illustrating Fe $K_{\alpha}$ and $K_{\beta}$ peaks in virgin and carbon-ion-irradiated HOPG samples.
.}}
\label{PIXE.jpg}
\end{figure}

In the present study, meticulous precautions were taken to avoid contamination during all stages of sample handling, including irradiation and magnetic measurements in the VSM system. To further confirm the absence of any transition-metal contamination, Particle Induced X-ray Emission (PIXE) analysis was carried out for both the as-received (virgin) HOPG sample and the irradiated sample after magnetic characterization, as shown in Figure \ref{PIXE.jpg} The PIXE spectra revealed detectable signals only for Fe $K_{\alpha}$ and $K_{\beta}$ lines in both samples, while no distinct peaks corresponding to other transition metals such as Co or Ni were observed.

Magnetization measurements performed before and after irradiation showed a clear enhancement in the magnetic moment following ion irradiation. This confirms that the observed magnetic enhancement arises due to irradiation-induced effects rather than impurity contributions, as any impurity-related magnetism would have been evident even in the unirradiated sample. The comparison of PIXE results indicates that the total iron impurity level in the irradiated sample remains below 20 ppm approximately twice the value reported by the supplier.

\section{Discussion}
Energetic protons and carbon ions interacting with HOPG during microbeam scanning transfer kinetic energy to target nuclei and electrons, producing point defects such as vacancies and Frenkel pairs \cite{telkhozhayeva2024}. Some defects recombine rapidly, while others persist, forming localized moments at dangling bonds or chemisorbed sites that mediate defect-induced ferromagnetic ordering.

In our study, proton irradiation (1–3 MeV) and carbon irradiation (0.6–2 MeV) revealed a strong energy dependence of magnetic ordering. SRIM simulations indicate that at lower energies, nuclear energy loss dominates, producing shallow defects, whereas higher energies increase electronic energy loss, leading to deeper ion penetration and a more dispersed defect profile. The optimal magnetic response occurs at intermediate energies, where the defect density in the magnetically active region is sufficient to promote long-range coupling without over-saturating the lattice.
Carbon ions generate a higher density of lattice defects compared to protons at the same fluence, consistent with the observed stronger magnetization. Proton irradiation additionally allows hydrogen chemisorption, forming quasi-localized $\pi$-electron states with higher local density of states near the Fermi level than vacancy defects, further contributing to magnetism\cite{Allemand1991}. The spatial extent of these states (3–10 nm) aligns with the scale over which magnetic ordering emerges \cite{Coey2002}.

These results demonstrate that both the ion species and beam energy critically govern the formation and distribution of defects, which in turn control the induced ferromagnetic behavior. Controlled microbeam irradiation thus provides a tunable route to engineer localized magnetic order in carbon-based materials.

\section{ Conclusions}
In this work, we have demonstrated that 1.2 MeV carbon microbeam scanning induces stable and robust room-temperature ferromagnetism in HOPG at an optimum fluence of $100~\text{pC}/\mu\text{m}^2$. Comparative analysis with proton irradiation reveals that carbon ions produce a more pronounced magnetic ordering, with the effect maximized within a well-defined threshold window of ion energy and fluence. Below and beyond this threshold, the magnetic ordering diminishes, underscoring the importance of controlled irradiation parameters. The maximum induced magnetic moment was found to be of the order of $1.0 \times 10^{-7}  \text{emu/cm}^3$ in fullerene thin films under 1.2 MeV carbon microbeam irradiation.

SRIM simulations supported these observations by showing that defect formation is shallow and localized at low energies, while deeper ion penetration at higher energies disperses energy, reducing effective magnetic coupling. Carbon ions generated greater lattice disruption than protons, directly correlating with higher magnetic moments.

Overall, our findings confirm that both ion energy and species decisively influence the defect-induced magnetic behavior of HOPG. The ability to engineer localized ferromagnetic ordering using microbeam scanning provides a promising pathway for developing patterned magnetic structures with potential applications in lightweight magnets, spintronic devices, and carbon-based memory technologies.

\section{Acknowledgments}
The authors are grateful for the support of the Tandetron Accelerator Laboratory, IIT Kanpur, for providing access to the microbeam facility. We also acknowledge the financial support from the IIT Kanpur through the Institute Postdoctoral Fellowship (IPDF), Grant No. PDF473. Additionally, we thank Mr. M. Siva Kumar for his technical assistance with VSM measurements at the Advanced Center for Materials Sciences, IIT Kanpur.

\bibliography{apssamp} 

@article{Stefania2022,
title = {Effects of Cs$^{+}$ and Ar$^{+}$ ion bombardment on the damage of graphite crystals},
journal = {Applied Surface Science},
volume = {585},
pages = {152756},
year = {2022},
issn = {0169-4332},
doi = {https://doi.org/10.1016/j.apsusc.2022.152756},
url = {https://www.sciencedirect.com/science/article/pii/S016943322200335X},
author = {Stefania {De Rosa} and Paolo Branchini and Valentina Spampinato and Alexis Franquet and Gianlorenzo Bussetti and Luca Tortora}
}

@article{KUMAR2023a,
title = {Strain induced structural changes and magnetic ordering in thin MoS$_{2}$ flakes as a consequence of 1.5 MeV proton ion irradiation},
journal = {Journal of Alloys and Compounds},
volume = {951},
pages = {169882},
year = {2023},
issn = {0925-8388},
doi = {https://doi.org/10.1016/j.jallcom.2023.169882},
url = {https://www.sciencedirect.com/science/article/pii/S0925838823011854},
author = {Ram Kumar and Aditya H. Kelkar and Rahul Singhal and Vasant G. Sathe and Ram Janay Choudhary and Neeraj Shukla}
}

@article{Ziegler2010,
title = {SRIM – The stopping and range of ions in matter (2010)},
journal = {Nuclear Instruments and Methods in Physics Research Section B: Beam Interactions with Materials and Atoms},
volume = {268},
number = {11},
pages = {1818-1823},
year = {2010},
note = {19th International Conference on Ion Beam Analysis},
issn = {0168-583X},
doi = {https://doi.org/10.1016/j.nimb.2010.02.091},
url = {https://www.sciencedirect.com/science/article/pii/S0168583X10001862},
author = {James F. Ziegler and M.D. Ziegler and J.P. Biersack}
}

@book{Ziegler1985,
author="Ziegler, James F.
and Biersack, Jochen P.",
editor="Bromley, D. Allan",
title="The Stopping and Range of Ions in Matter",
bookTitle="Treatise on Heavy-Ion Science: Volume 6: Astrophysics, Chemistry, and Condensed Matter",
year="1985",
publisher="Springer US",
address="Boston, MA",
pages="93--129",
isbn="978-1-4615-8103-1"
}

@article{Allemand1991,
author = {Pierre-Marc Allemand  and Kishan C. Khemani  and Andrew Koch  and Fred Wudl  and Karoly Holczer  and Steven Donovan  and George Grüner  and Joe D. Thompson },
title = {Organic Molecular Soft Ferromagnetism in a Fullerene C$_{60}$},
journal = {Science},
volume = {253},
number = {5017},
pages = {301-302},
year = {1991},
doi = {10.1126/science.253.5017.301},
URL = {https://www.science.org/doi/abs/10.1126/science.253.5017.301}
}

@article{Mathew2012,
   author = {S. Mathew and K. Gopinadhan and T. K. Chan and X. J. Yu and D. Zhan and L. Cao and A. Rusydi and M. B.H. Breese and S. Dhar and Z. X. Shen and T. Venkatesan and John T.L. Thong},
   doi = {10.1063/1.4750237},
   issn = {00036951},
   issue = {10},
   journal = {Applied Physics Letters},
   title = {Magnetism in MoS$_{2}$ induced by proton irradiation},
   volume = {101},
   year = {2012},
}

@article{Kumar2022b,
   author = {Ram Kumar and Neeraj Shukla},
   doi = {10.1007/s00339-022-06254-w},
   issn = {0947-8396},
   issue = {12},
   journal = {Applied Physics A},
   pages = {1096},
   title = {Weak ferromagnetism in thin fullerene films as a consequence of 600 keV carbon ion irradiation},
   volume = {128},
   url = {https://link.springer.com/10.1007/s00339-022-06254-w},
   year = {2022},
}

@article{Kumar2022a,
   author = {R. Kumar and K. Mohan and A. Augusthy and S. Bari and A.P. Parhi and A.H. Kelkar and S. Chakravarty and N. Shukla},
   doi = {10.1016/j.tsf.2022.139350},
   issn = {00406090},
   journal = {Thin Solid Films},
   title = {Tunable room temperature ferromagnetism in fullerene thin film induced by 1 MeV proton microbeam irradiation},
   volume = {755},
   year = {2022},
}

@article{Makarova2011,
   author = {Tatiana L. Makarova and Oleg E. Kvyatkovskii and Irina B. Zakharova and Sergei G. Buga and Aleksandr P. Volkov and Andrei L. Shelankov},
   doi = {10.1063/1.3581105},
   issn = {00218979},
   issue = {8},
   journal = {Journal of Applied Physics},
   title = {Laser controlled magnetism in hydrogenated fullerene films},
   volume = {109},
   year = {2011},
}

@article{Lungu2021,
   author = {Iulia Ioana Lungu and Alexandru Mihai Grumezescu and Claudiu Fleaca},
   doi = {10.3390/app11156707},
   issn = {20763417},
   issue = {15},
   journal = {Applied Sciences (Switzerland)},
   publisher = {MDPI AG},
   title = {Unexpected ferromagnetism - A review},
   volume = {11},
   year = {2021},
}

@article{Esquinazi2002,
   author = {P. Esquinazi and A. Setzer and R. Höhne and C. Semmelhack and Y. Kopelevich and D. Spemann and T. Butz and B. Kohlstrunk and M. Lösche},
   doi = {10.1103/PhysRevB.66.024429},
   issn = {1550235X},
   issue = {2},
   journal = {Physical Review B - Condensed Matter and Materials Physics},
   pages = {1-10},
   title = {Ferromagnetism in oriented graphite samples},
   volume = {66},
   year = {2002},
}

@article{Wang2014,
   author = {Yutian Wang and Pascal Pochet and Catherine A. Jenkins and Elke Arenholz and Gregor Bukalis and Sibylle Gemming and Manfred Helm and Shengqiang Zhou},
   doi = {10.1103/PhysRevB.90.214435},
   issn = {1550235X},
   issue = {21},
   journal = {Physical Review B - Condensed Matter and Materials Physics},
   publisher = {American Physical Society},
   title = {Defect-induced magnetism in graphite through neutron irradiation},
   volume = {90},
   year = {2014},
}

@article{Coey2002,
   author = {J. M.D. Coey and M. Venkatesan and C. B. Fitzgerald and A. P. Douvalis and I. S. Sanders},
   doi = {10.1038/nature01100},
   issn = {00280836},
   issue = {6912},
   journal = {Nature},
   pages = {156-159},
   pmid = {12432386},
   title = {Ferromagnetism of a graphite nodule from the Canyon Diablo meteorite},
   volume = {420},
   year = {2002},
}

@article{Shukla2012,
   author = {Neeraj Shukla and Mihir Sarkar and Nobin Banerji and Anjan K. Gupta and Harish C. Verma},
   doi = {10.1016/j.carbon.2011.12.031},
   issn = {00086223},
   issue = {5},
   journal = {Carbon},
   pages = {1817-1822},
   title = {Inducing large ferromagnetic ordering in graphite by 1 MeV Carbon ion irradiation},
   volume = {50},
   year = {2012},
}

@article{Talapatra2006,
   author = {S. Talapatra and Ju Yin Cheng and N. Chakrapani and S. Trasobares and A. Cao and R. Vajtai and M. B. Huang and P. M. Ajayan},
   doi = {10.1088/0957-4484/17/1/052},
   issn = {09574484},
   issue = {1},
   journal = {Nanotechnology},
   pages = {305-309},
   title = {Ion irradiation induced structural modifications in diamond nanoparticles},
   volume = {17},
   year = {2006},
}

@article{Kumar2006,
   author = {Amit Kumar and D. K. Avasthi and J. C. Pivin and A. Tripathi and F. Singh},
   doi = {10.1103/PhysRevB.74.153409},
   issn = {10980121},
   issue = {15},
   journal = {Physical Review B - Condensed Matter and Materials Physics},
   title = {Ferromagnetism induced by heavy-ion irradiation in fullerene films},
   volume = {74},
   year = {2006},
}

@article{Makarova2010,
   author = {Tatiana L. Makarova and Andrei L. Shelankov and Igor T. Serenkov and Vladimir I. Sakharov},
   doi = {10.1002/pssb.201000114},
   issn = {03701972},
   issue = {11-12},
   journal = {Physica Status Solidi (B) Basic Research},
   pages = {2988-2991},
   title = {Magnetism in graphite induced by irradiation of hydrogen or helium ions - A comparative study},
   volume = {247},
   year = {2010},
}

@article{Mathew2007,
   author = {S. Mathew and B. Satpati and B. Joseph and B. N. Dev and R. Nirmala and S. K. Malik and R. Kesavamoorthy},
   doi = {10.1103/PhysRevB.75.075426},
   issn = {10980121},
   issue = {7},
   journal = {Physical Review B - Condensed Matter and Materials Physics},
   title = {Magnetism in C60 films induced by proton irradiation},
   volume = {75},
   year = {2007},
}

@article{Esquinazi2003,
   author = {P. Esquinazi and D. Spemann and R. Höhne and A. Setzer and K. H. Han and T. Butz},
   doi = {10.1103/PhysRevLett.91.227201},
   issn = {10797114},
   issue = {22},
   journal = {Physical Review Letters},
   title = {Induced magnetic ordering by proton irradiation in graphite},
   volume = {91},
   year = {2003},
}

@article{WANG2022,
title = {The surface defects of HOPG induced by low-energy Ar+ ion irradiation},
journal = {Applied Surface Science},
volume = {585},
pages = {152680},
year = {2022},
issn = {0169-4332},
doi = {https://doi.org/10.1016/j.apsusc.2022.152680},
url = {https://www.sciencedirect.com/science/article/pii/S0169433222002616},
author = {Xiaogang Wang and Guopeng Li and Luyao Zhang and Feifei Xiong and Yue Guo and Guang Zhong and Jiawei Wang and Pinyang Liu and Yuanqing Shi and Yanling Guo and Lin Chen and Ximeng Chen}
}

@article{telkhozhayeva2024,
  title={Roadmap toward controlled ion beam-induced defects in 2D materials},
  author={Telkhozhayeva, Madina and Girshevitz, Olga},
  journal={Advanced Functional Materials},
  volume={34},
  number={45},
  pages={2404615},
  year={2024},
  publisher={Wiley Online Library}
}

@article{walker2016,
  title={Radiation effects on two-dimensional materials},
  author={Walker, RC and Shi, T and Silva, EC and Jovanovic, I and Robinson, JA},
  journal={physica status solidi (a)},
  volume={213},
  number={12},
  pages={3065--3077},
  year={2016},
  publisher={Wiley Online Library}
}

\end{document}